\documentclass[superscriptaddress,nobibnotes,amsmath,amssymb,notitlepage,twocolumn,pra,longbibliography]{revtex4-2}
\pdfoutput=1

\usepackage{bm,braket}
\usepackage[toc,page]{appendix}
\usepackage{comment}
\usepackage{dcolumn}

\usepackage{amsfonts}

\usepackage{graphicx,color,hyperref}
\usepackage[caption=false]{subfig}
\hypersetup{colorlinks=true, linkcolor=blue, citecolor=blue, urlcolor=blue} 

\graphicspath{{./img/}}

\newcommand{\beq}[1]{\begin{equation}\label{#1}}
\newcommand{\eep}{\;.\end{equation}}
\newcommand{\eec}{\;,\end{equation}}
\newcommand{\eeq}{\end{equation}}

\newcommand*\dd{\mathop{}\!\mathrm{d}} 

\newcommand{\Om}{\Omega}

\DeclareMathAlphabet{\mathcal}{OMS}{cmsy}{m}{n} 

\renewcommand{\vec}[1]{{\bf #1}}

\newcommand{\kv}{\vec{k}}

\usepackage{amsmath}
\usepackage{amssymb}
\usepackage{xcolor}
\usepackage{bbm}
\usepackage{physics}
\usepackage{float}
\usepackage{graphicx}
\usepackage{dcolumn} 
\usepackage{bm} 
\usepackage{siunitx}
\usepackage{enumitem}  

\makeatletter
\renewcommand*{\fnum@figure}{{\normalfont\bfseries \figurename~\thefigure}}
\makeatother

\allowdisplaybreaks

\definecolor{orange}{rgb}{1,0.5,0}

\newcommand{\sect}[1]{\vspace{0.3em}{\it #1.}---}

\graphicspath{{./img/}}

\DeclareMathAlphabet{\mathcal}{OMS}{cmsy}{m}{n} 

\newcommand{\intBZ}{\int_{\text{BZ}}} 

\makeatletter
\newcommand{\specificthanks}[1]{\@fnsymbol{#1}}
\makeatother

\begin{document}

\preprint{APS/12three-QED}

\title{Nonlinear Odd Viscoelastic Effect}

\newcommand{\TCM}{{Theory of Condensed Matter Group, Cavendish Laboratory, University of Cambridge, J.\,J.\,Thomson Avenue, Cambridge CB3 0HE, UK}}
\newcommand{\UoM}{Department of Physics and Astronomy, University of Manchester, Oxford Road, Manchester M13 9PL, UK}
\newcommand{\KITP}{Kavli Institute for Theoretical Physics, University of California, Santa Barbara, CA 93106, USA}


\author{Ashwat Jain}
\email{ashwat.jain@postgrad.manchester.ac.uk}
\affiliation{\UoM}

\author{Wojciech J. Jankowski}
\email{wjj25@cam.ac.uk}
\affiliation{\TCM}
\affiliation{\KITP}

\author{M. Mehraeen}
\email{mandela.mehraeen@manchester.ac.uk}
\affiliation{\UoM}

\author{Robert-Jan Slager}
\email{robert-jan.slager@manchester.ac.uk}
\affiliation{\UoM}
\affiliation{\TCM}

\date{\today}

\begin{abstract}
    We uncover a class of nonlinear odd viscoelastic effects in three spatial dimensions. We show that these dissipationless effects arise upon combining geometric distortions in two orthogonal directions, yielding momentum flow in the third direction. We demonstrate that the effect arises from nontrivial geometric tensors in quantum states, and can be scaled up with integer topological invariants. We further show that the effect fingerprints the multiband Hilbert-space geometry of the underlying quantum states, as encoded in the nonmetricity and three-state quantum geometric tensors. Our findings unravel the role of multistate geometry in viscoelastic phenomena, paving a path for experimental observation of uncharted nonlinear odd viscoelastic responses in quantum systems.
\end{abstract}

\maketitle

\sect{Introduction}Viscosity and elasticity are bulk properties of matter that manifest upon application of physical deformations. They naturally arise in a wide range of physical systems at different scales spanning from \mbox{cosmology} and astrophysics~\cite{Padmanabhan1987, Brevik2017}, to particle physics~\cite{Policastro2001}, through mesoscopic fluids~\cite{Monaghan2012}, glasses~\cite{Dyre2006}, and active condensed matter systems~\cite{Ramaswamy2010}, among others. The corresponding transport coefficients that determine the responses to geometric perturbations imposed by normal and shear stresses are encoded in viscoelastic tensors, which are in turn captured by stress correlators~\cite{LL6,LL7,TongKin}. Beyond their distinct emergence in classical fluids and solids, viscoelastic tensors are also realized by quantum fluids of electrons in~crystals. In particular, viscoelastic tensors also include dissipationless flows of momentum orthogonal to perturbation directions, commonly known as odd viscosity and odd elasticity~\cite{Banerjee2017, Offertaler2019, Scheibner2020, Fruchart2023, Chen2024}. In condensed matter systems, odd viscosity arises in viscous Hall currents, that is, as Hall viscosity~\cite{Avron1995, Vignale2007, Hoyos2012, Bradlyn2012, Gromov2015, Rao2020}. Odd viscosity has been experimentally measured in crystals such as graphene~\cite{Berdyugin2019}, and was further recognized to contribute corrections to spatially dispersive conductivities~\cite{Hoyos2012, Sherafati2016}.

Hall phenomena have, among many other impactful directions, been related to two themes of study. First, there is a primary role for topology, well exemplified by various detailed responses of quantum Hall fluids~\cite{Laughlin1983, Niu1985, vKlitzing1986, Stormer1999, Yu2010, Read2011}. Second, it has been demonstrated that intrinsic electronic Hall effects occur not only at linear order in perturbing electric fields~\cite{nagaosa2010anomalous}, but are also manifested at higher orders~\cite{gao2014field, Sodemann2015, wang2021intrinsic, Jankowski2024PRL, liu2021intrinsic, Wang2023a, Gao2023, Mehraeen2025,guo2025bicircular}. In addition, electronic nonlinear Hall responses are an increasingly studied topic of recent interest in quantum Hall fluids and topological phases, from both experimental and theoretical angles~\cite{ortix2021nonlinear, ideue2021symmetry, du2021nonlinear, zhang2022large, Kaplan2023, mehraeen2024proximity, shim2025spin, suarez2025nonlinear, Zhang2022, Jain2025}. Universally, these responses depend heavily on the Hilbert-space geometry of quantum states, which is intricately linked to, and can be controlled by, the topology of wave~functions~\cite{Klevtsov2015, Xie2020, Ahn2020, Ahn2021, Bouhon2023, Oancea2025}. 
 
Given the well-established understanding of these fields and the rich quantum geometric structure of nonlinear electromagnetic responses beyond the extensively studied Berry curvature and quantum metric tensors, a natural question emerges as to whether new insights can emerge upon exploiting the relation between multiband geometry, topology, higher-order responses, and 
odd viscoelasticity. Notably, a nonlinear viscoelastic effect that might accompany, or more importantly, arise even in the absence of, electronic Hall currents has to our knowledge not been explored to date, particularly within these quantum geometric frameworks.
\begin{figure}
    \centering
    \includegraphics[width=\linewidth]{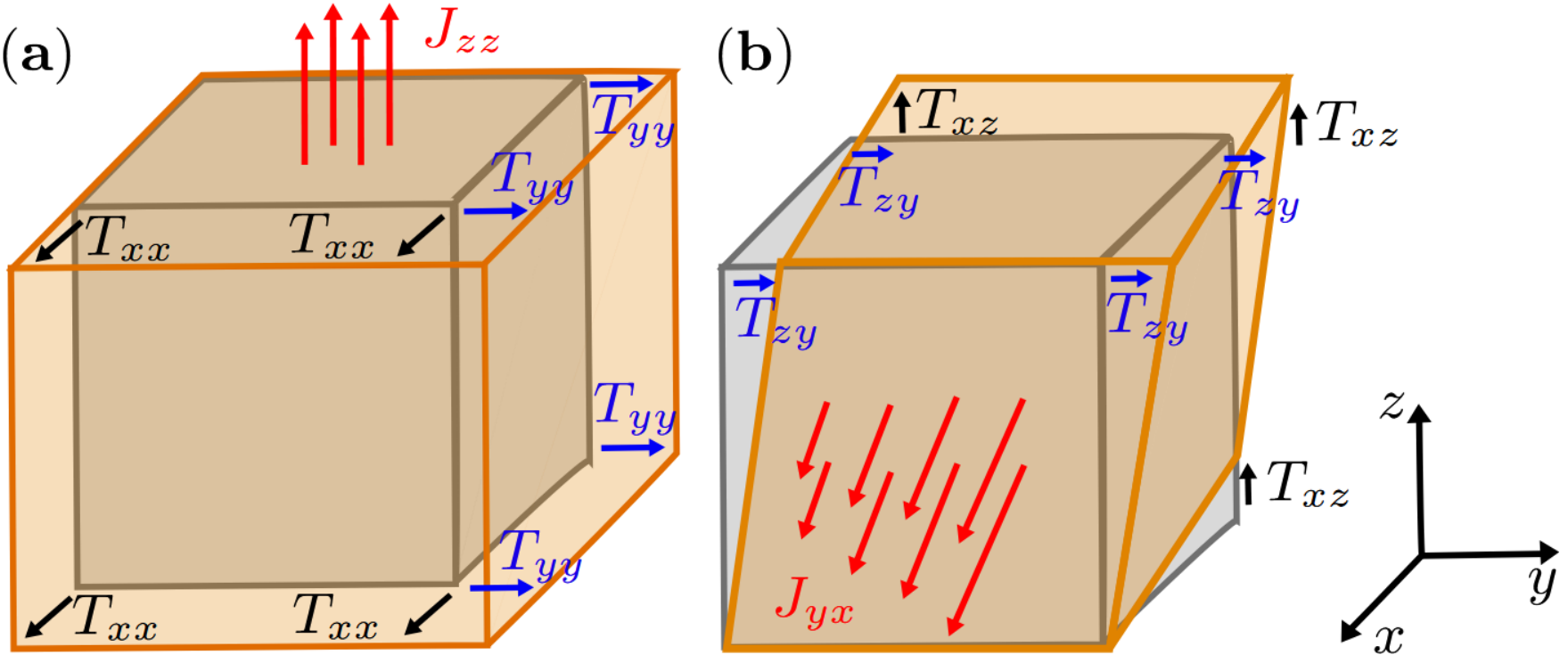}
    \caption{Nonlinear odd viscoelastic effects. Application~of {\bf (a)} orthogonal normal distortions $w_{xx}, w_{yy}$ coupling to stress-momentum currents $T_{xx}, T_{yy}$, and {\bf (b)} shear distortions $w_{xz}, w_{zy}$ coupling to currents $T_{xz}, T_{zy}$, result in dissipationless nonlinear viscoelastic current responses ${J_{zz} = \langle T_{zz}\rangle}$ and ${J_{yx} = \langle T_{yx} \rangle}$ along the normal to the stress plane, respectively.}
    \label{fig:schematics}
\end{figure}
\begin{figure}
    \centering
    \includegraphics[width=\linewidth]{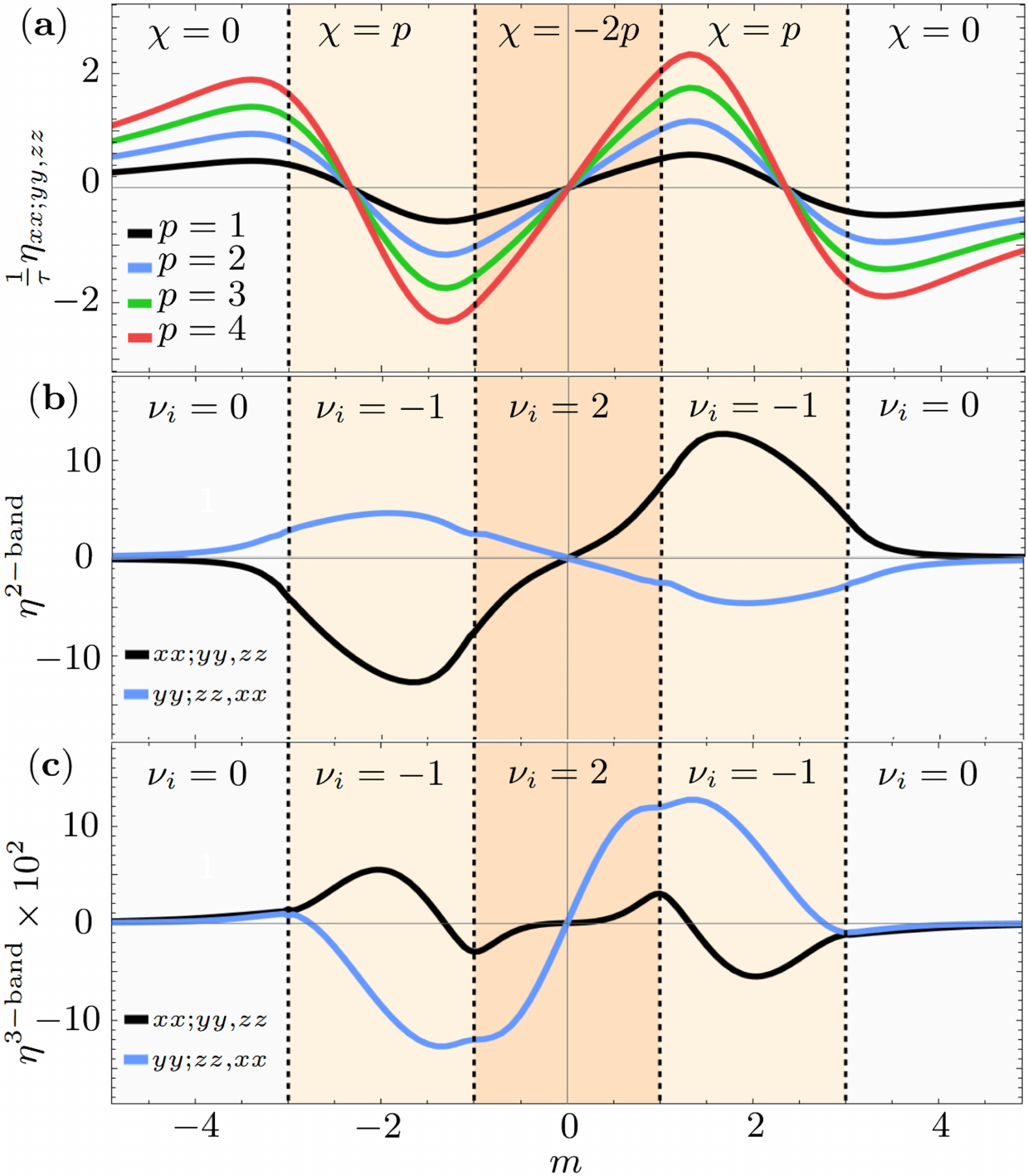}
    \caption{Two- and three-band contributions to the nonlinear odd elasticity. {\bf (a)} Two-band $\eta_{xx;yy,zz}$ for the MRW model with mass $m$, showcasing a sustained response in the phases with nonzero Hopf invariant magnitudes $|\chi|$ and a decaying response in the trivial phase $(\chi=0)$. Here, $p \in \mathbb{Z}$ denotes the momentum scaling factor $k_z \rightarrow p k_z$ which leads to integer scaling of the invariant and the response. {\bf (b)} Two-band and {\bf (c)} three-band $\eta_{xx;yy,zz}$ and $\eta_{yy;zz,xx}$ realized in a flattened three-dimensional chiral model with chiral invariant $|\nu_i|$ that trivializes upon adding a symmetry-breaking perturbation. The $\eta_{zz;xx,yy}$ response component follows from the sum rule $\eta_{xx;yy,zz}+\eta_{yy;zz,xx}+\eta_{zz;xx,yy}=0$.}
    \label{fig:Hall}
\end{figure}

In this Letter, we predict the emergence of such an effect and, in doing so, also address the problem of the existence of nonlinear \textit{odd} viscoelastic effects (NOVEs) in quantum fluids. We demonstrate that the nonlinear correlators of elastic stress and modulus operators are uniquely constituted by two-state and three-state quantum geometric tensors (QGTs), revealing the crucial role of multistate geometry in viscoelastic responses. We also show that the quantum nonmetricity tensor and three-state QGT can be isolated via NOVEs in the flat-band limit, and~that two- and three-state geometric responses arise naturally in magnetic topological phases, establishing these systems as strong candidates for measuring viscoelastic nonlinearities. 

Moreover, we demonstrate the interplay of nonlinear elasticity and ground-state topologies via qualitative changes of nonlinear stress correlators, culminating in marked variations of NOVEs across topological phase transitions. Finally, we showcase the nonlinear viscoelastic phenomenology via deformations of real-space Wannier functions, subject to nonlinear couplings to normal and shear stresses. Our findings establish a set of unique viscoelastic effects and experimental protocols to measure them in a broad range of phases of matter. Beyond allowing to access and fingerprint exotic magnetic topological invariants, our results establish viscoelastic responses as probes of nontrivial multistate geometries for electronic states in solids.\\

\sect{Nonlinear odd viscoelastic response}To retrieve the NOVE, presented schematically in Fig.~\ref{fig:schematics}, we consider the nonlinear response of the system to the elastic distortion fields $w_{ij} = \partial \mathfrak{u}_i /\partial {x_j}$, with $\boldsymbol{\mathfrak{u}}$ denoting local displacements in the system, $\vec{x} \rightarrow \vec{x} + \boldsymbol{\mathfrak{u}}$.


In a crystalline solid, the geometric deformations of the system can be captured by minimally coupling the distortion fields to momenta in the lattice Hamiltonian as
$H(k_j)
\rightarrow
H[ k_j - w_{ij}\frac{\sin(k_i a_0)}{a_0}]$, with $a_0$ the lattice constant~\cite{Shapourian2015}. At arbitrary frequencies $\omega_{1}, \omega_{2}$ of the applied stress perturbations, the nonlinear elastic current density takes the general form
\begin{equation}
    J_{ij} (\omega)
    =
    \eta_{ij;kl,mn} (\omega; \omega_1, \omega_2)
    w_{kl}(\omega_1) w_{mn} (\omega_2),
\end{equation}
with $\{i,j,k,l,m,n\} = \{x,y,z \}$. To evaluate the nonlinear elastic response tensor $\eta_{ij;kl,mn}$ in the dc limit ${\omega_1, \omega_2 \rightarrow 0}$, we employ quadratic Kubo response theory generalized to higher-order tensor distortion field perturbations. In the systematic perturbative expansion, in addition to the stress tensor operator $T_{ij} = \delta H/\delta w_{ij}$, we identify higher-order operators, i.e., the elastic modulus tensor operator
$U_{ij,kl} = \delta^2 H/\delta w_{ij} \delta w_{kl}$ and the nonlinear elastic modulus tensor operator
$V_{ij,kl,mn} = \delta^3 H/\delta w_{ij} \delta w_{kl} \delta w_{mn}$, to be essential for a comprehensive derivation of nonlinear viscoelastic responses. The tensor $U$ encodes a viscoelastic analog of diamagnetic response. The tensor $V$ captures the contact terms in quadratic response theory, although these contributions vanish for odd responses. We express $T$,~$U$, and $V$ in terms of band-geometric quantities in the End Matter, following derivations in Supplemental Material (SM)~\cite{SI}.

The evaluation of the second-order correlators for odd viscoelastic responses reveals a rich quantum geometric structure. Specifically, we find that the dc quadratic response tensor naturally decomposes into gauge-invariant two-state and three-state geometric terms as
\begin{equation}
\label{eq:final_visc_sigma}
   \eta_{ij;kl,mn} = \int_{\mathrm{BZ}} \mathrm{d}^3\vec{k}~ \big[\eta^\mathrm{2\text{-band}}_{ij;kl,mn}(\vec{k}) + \eta^\mathrm{3\text{-band}}_{ij;kl,mn}(\vec{k})\big].
\end{equation}
Here, we integrate over lattice momenta $\textbf{k}$ in the three-dimensional Brillouin zone (BZ). The first term, $\eta^\mathrm{2\text{-band}}_{ij;kl,mn}(\vec{k})$, captures the two-band contributions to viscoelastic effects at zero temperature in systems with a finite energy gap, such as trivial and topological insulators. This measures the response arising from the quantum state geometry of virtual transitions of occupied states to the unoccupied state manifold induced by the stress tensor and elastic moduli operators. The second term, $\eta^\mathrm{3\text{-band}}_{ij;kl,mn}(\vec{k})$, arises from the three-band contributions, which emerge from (i) triplets of virtual transitions between an occupied state and two virtual unoccupied energy levels and (ii) additional contributions with two occupied states and one empty virtual energy level~\cite{SI}.

We further characterize the quantum geometries \mbox{underlying} the individual NOVE terms. The two-band contributions, resolved in momentum space, compactly read
\begin{align}
\label{eq:2band}
   \eta^\mathrm{2\text{-band}}_{ij;kl,mn}(\vec{k})
   =& 
   \sum_{a,b}f_{ba}\bigg[ \frac{4\text{i} \tau}{3} g^{ba}_{[(kl,mn)}\partial_{ij]}\Delta^{ba} \nonumber
   \\
   &+ \frac{\Omega^{ba}_{ij,(kl}\partial_{mn)}\Delta^{ba}}{\Delta^{ba}} + \frac{32\text{i}}{3}N^{ba}_{[(kl,mn),ij]}\bigg],
\end{align}
where $\partial_{kl} \equiv \partial/\partial w_{kl}$ are distortion derivatives, ${\Delta^{ba} \equiv E^{b}_{\kv} - E^{a}_{\kv}}$ are eigenstate energy differences, ${f_{ba} = f_b - f_a}$ is the difference of band occupation factors, and $\tau$ is the scattering time that enters as the inverse of the Lorentzian broadening in the electron propagators in the dc limit. The adiabatic curvature, $\Omega_{ij,kl}^{ba} = -2~\text{Im}(A_{ij}^{ba}A_{kl}^{ab})$, is a two-form in the parameter space of deformations, $g^{ba}_{ij,kl} = \text{Re}(A_{ij}^{ba}A_{kl}^{ab})$ is the corresponding metric, $N^{ba}_{ij,kl,mn} = -\Tilde{\nabla}_{mn}g^{ba}_{ij,kl}$ is the quantum nonmetricity tensor (see End Matter). $(\dots)$ and $[\ldots]$ denote normalized symmetrizations and antisymmetrizations of spatial index pairs, respectively. Here, $A_{ij}^{ab} = \text{i}\bra{u^{a}_{\kv}}\ket{\partial_{ij} u^{b}_{\kv}} = \frac{-\sin(k_i a_0)}{a_0} A_j^{ab}$ is the non-Abelian Berry connection in the parameter space of deformations, with $A_j^{ab}$ the momentum-space Berry connection. 

Physically, the quantum nonmetricity tensor $N$ captures the failure of the Hermitian connection to preserve the two-state quantum metric under parallel transport. Equivalently, it quantifies the covariant variation of the matrix-valued inner product between quantum-state bundle vectors along momentum-space geodesics.
 
Additionally, the three-band contributions to the dc nonlinear elastic response tensor are concisely expressed as
\begin{align}
\label{eq:3band}
   \eta^\mathrm{3\text{-band}}_{ij;kl,mn}(\vec{k})
   =&
   \frac{2}{3} \sum\limits_{a,b,c} \frac{\text{Re}~Q_{ij,(kl,mn)}^{abc}}{\Delta^{ba}}\bigg(2+\frac{(\Delta^{ba})^2}{\Delta^{bc}\Delta^{ca}}\bigg)\nonumber
   \\
   & \times (f_a \Delta^{bc}+f_b \Delta^{ca}+f_c \Delta^{ab}),
\end{align}
with $Q_{ij,kl,mn}^{abc}=A_{ij}^{ab} A_{kl}^{bc} A_{mn}^{ca}$ a three-state QGT~\cite{Mehraeen2025, guo2025bicircular} in the deformation parameter space. The three-state QGT captures the amplitudes of interband transitions between three states~\cite{Ahn2021, Jankowski2024PRL, Mehraeen2025}, which reflects the nonlinear character of the odd viscoelastic effect. For full derivations of $\eta^\mathrm{2\text{-band}}_{ij;kl,mn}(\vec{k})$, $\eta^\mathrm{3\text{-band}}_{ij;kl,mn}(\vec{k})$ and a detailed analysis of the symmetry transformations and of the independent components of the response tensors, see~the~SM~\cite{SI}.

\sect{Example I: Nonlinear viscoelasticity of Hopf bands}We now turn to concrete model settings to exemplify the controllable effect in two topological bands. The NOVEs under consideration require inherently three-dimensional quantum geometry, which mixes all three spatial dimensions. Thus, we appeal to basic three-dimensional models of magnetic topological insulators with nontrivial topologies, which naturally provide nontrivial quantum geometries~\cite{Jankowski2025gerbe}. To this end, consider a two-band Moore-Ran-Wen (MRW) Hopf insulator model~\cite{Moore2008}, see End Matter. The integer topological Hopf invariant, $\chi \in \mathbb{Z}$, reads: ${\chi = -\frac{1}{4\pi^2} \intBZ \dd^3 \kv~\varepsilon^{ijk} A_i^{aa} \Om_{jk}^{aa}}$, with ${\Om_{jk}^{aa} = \partial_{k_j} A_k^{aa} - \partial_{k_k} A_j^{aa}}$ the band-diagonal Berry curvature tensor. The invariant requires nontrivial Berry curvature $\Omega_{jl}$, and by extension, $\Omega_{ij,kl}$ tensors, and thus serves as an effective source of nontrivial quantum geometry required for $\eta^\mathrm{2\text{-band}}_{ij;kl,mn}$, as we demonstrate in Fig.~\ref{fig:Hall}(a).  Here, we can access an arbitrarily high Hopf invariant $\chi \rightarrow p \chi$ upon scaling $k_z \rightarrow p k_z$, with $p \in \mathbb{Z}$~\cite{Alexandradinata2021}. In addition, we may control topological phases across topological phase transitions by varying a topological mass parameter $m$. In Fig.~\ref{fig:Hall}(a), we show the changes and the integer $(p)$ scaling of the $\eta_{xx;yy,zz}$ nonlinear viscoelastic tensor component. Because of symmetry-enforced selection rules, under the combined momentum-space mirror and time reversal symmetry $\mathcal{M}_{xz}\mathcal{T}$, this model selectively realizes the first term in Eq.~\eqref{eq:2band}~\cite{SI}. Notably, the trivial phase $\chi=0$ ($|m|>3$) exhibits decaying odd viscoelastic effects, unlike its topological counterparts. We find that the response scales exactly with $p$ in the model, i.e., $\int_{\text{BZ}} \mathrm{d}^3\vec{k}~\eta(k_x, k_y, p k_z) = p \int_{\mathrm{BZ}} \eta(k_x, k_y, k_z)$~\cite{SI}.

\sect{Example II: Multiband nonlinear viscoelasticity}We~now focus on the demonstration of two- and three-band contributions to the elasticity in multiband systems, and highlight deeper connections of NOVEs to internal geometry. As a starting point, we adapt chiral three-dimensional models in the Altland-Zirnbauer class AIII~\cite{Kitaev2009}. We start with flattened perturbed three-band systems characterized by a chiral invariant, $\nu_i \in \mathbb{Z}$, ~\cite{Neupert2012,Palumbo2019}, see End Matter. The flat-band limit allows to isolate and probe $\eta^\mathrm{}_{ij;kl,mn}$ contributions from the quantum nonmetricity tensor $\sum_{a,b}f_{ba}N^{ba}_{[(kl,mn),ij]}$ in Eq.~\eqref{eq:2band}, and the three-band responses due to the three-state QGTs, $Q^{abc}_{ij,(kl,mn)}$, in Eq.~\eqref{eq:3band}. In this limit, the $\vec{k}$~dependence of both two- and three-band responses becomes entirely geometric and deconvolved from the band dispersion contributions, showcasing that NOVEs \mbox{directly} reflect the internal geometry of nonmetricities and three-state QGTs. Further retaining the dispersionless limit, we break the $\mathcal{M}_{xz}\mathcal{T}$ symmetry of the model, which otherwise causes the integrated $N$- and $Q$-dependent responses to vanish as a result of the aforementioned selection rules. This symmetry-breaking perturbation also trivializes the invariant, and demonstrates unambiguously that the retrieved effect is intrinsically geometric in nature, and goes beyond nontrivial topologies per se. We present numerical results for the two- and three-band contributions in this flattened perturbed chiral model in Fig.~\ref{fig:Hall}(b) and Fig.~\ref{fig:Hall}(c), respectively.

We observe that the responses are significantly weaker in the phase $|m| > 3$, owing to the suppression of the geometric tensors by the virtual transition matrix elements. In the considered model, this suppression further results in a $2$~order of magnitude difference between the two- and three-band contributions. While this chiral setup effectively demonstrates the geometric nature of the effect, we note in passing that the three-band effect can also be retrieved in models with a~preserved invariant, analogously to the two-band Hopf case. Indeed, when all bands are gapped and characterized by a classifying space $\mathsf{U}(3)/\mathsf{U}(1)^3$, this allows for an integer invariant (${\pi_3[\mathsf{U}(3)/\mathsf{U}(1)^3]=\mathbb{Z}}$) that does not rely on additional symmetries beyond translational symmetry and particle number conservation~\cite{Lapierre2021, Jankowski2025gerbe}.

\sect{Nonlinear viscoelasticity in Wannier basis}Having retrieved the two- and three-band viscoelastic momentum-space contributions in topological and geometrically nontrivial phases, we now characterize NOVEs in the real-space formalism. Correspondingly, we demonstrate how the nonlinear viscoelastic effects are manifested in the real-space Wannier function representation. To this end, we transform the single-particle states to the localized Wannier basis ${\ket{a\textbf{R}} = \intBZ \frac{\dd^3 \textbf{k}}{(2\pi)^3}~e^{-\text{i} \textbf{k} \cdot \textbf{R}} \ket{\psi^{a}_{\textbf{k}}}}$~\cite{Marzari1997, Marzari2012}, and deduce their centers of mass $x_c \equiv \bra{{a\textbf{0}}} \hat{x} \ket{{a\textbf{0}}}$ for the lowest band $a=1$, subject to perturbations $w_{kl}, w_{mn}$.
\begin{figure}[t!]
    \centering
    \includegraphics[width=\linewidth]{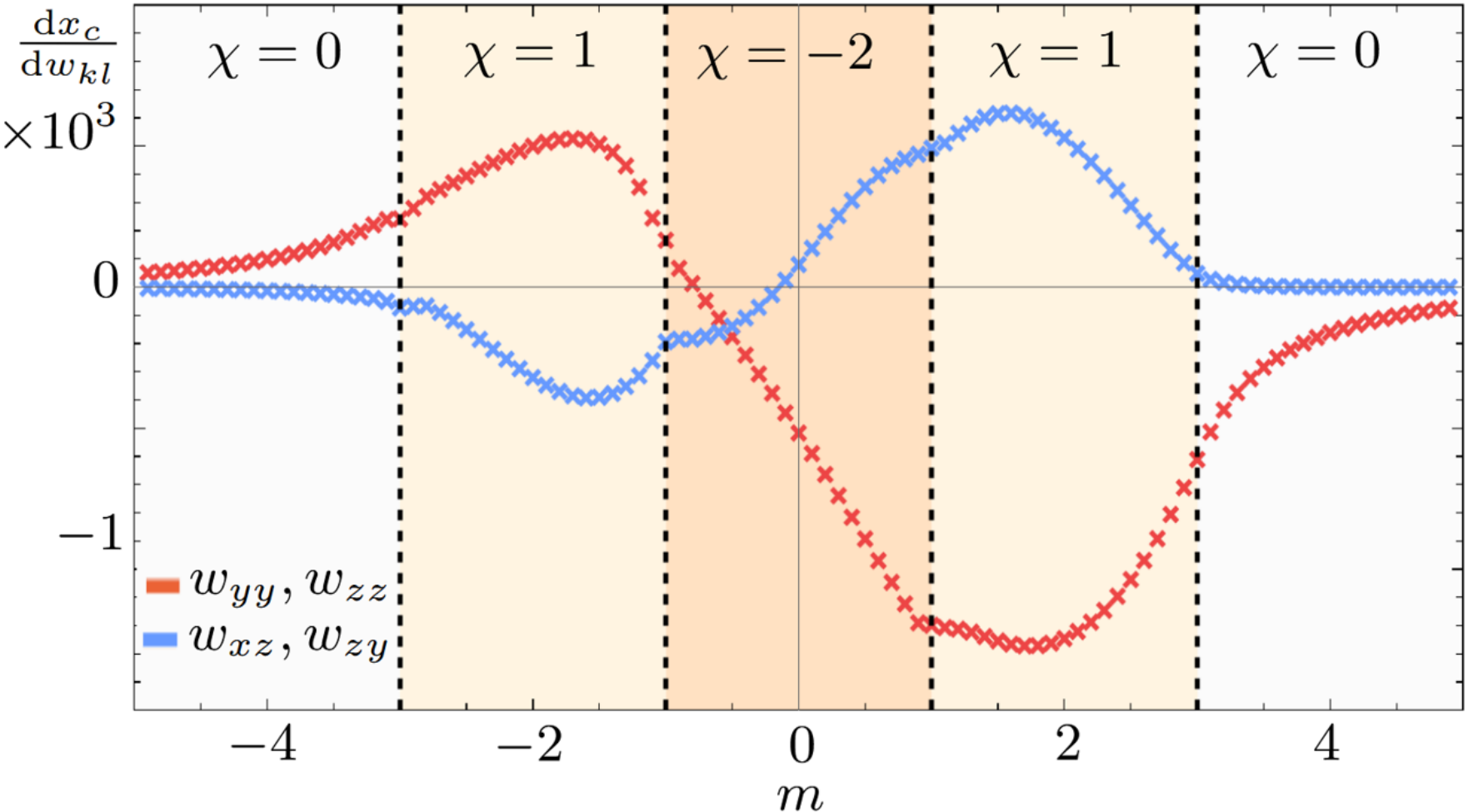}
    \caption{Real-space representation of the NOVE in the MRW model for $p = 1$. As a function of the topological mass $m$, we show the dependence of the ratio $x_c/w_{kl}$ of the Wannier center $x$~velocity $v_c = d x_c/ dt$ to the distortion rate $d w_{kl} / dt$ that induces it. We consider here equal applied normal stresses $w_{yy},w_{zz}$ (\text{red}) and equal shear stresses $w_{xz},w_{zy}$ (\text{blue}) that lead to momentum currents $J_{xx}$ and $J_{yx}$ in the $x$~direction. The viscoelastic response of the Wannier functions changes across topological phase transitions, displaying correspondence to the momentum-space results shown in Fig.~\ref{fig:Hall}.}
    \label{fig:Wannier}
\end{figure}

In Fig.~\ref{fig:Wannier}, we show the normal  and shear stress induced Wannier velocities $\dd x_c/\dd w_{kl}$ in different topological phases, which are consistent with the elastic response currents $J_{ij} = \langle T_{ij} \rangle$ shown in Fig.~\ref{fig:schematics}. Hence, we show that the localized wave~packet basis reflects the nonlinear odd viscoelastic transport induced by the deformations, consistently with the values of nonlinear response functions $\eta_{ij;kl,mn}$ demonstrated in Fig.~\ref{fig:Hall}. As such, we show that NOVEs naturally arise in momentum-space and real-space representations of fermionic wave~functions realizing three-dimensional quantum-state \mbox{geometries}.\\

\sect{Discussion and conclusion}We first highlight and discuss characteristic properties of NOVEs, which lie in the symmetry transformations of the nonlinear viscoelastic response tensors, the unique roles of quantum geometry underpinned by the Hilbert-space nonmetricity and three-state QGTs, and their dissipationless nature.

The symmetry transformations of NOVEs account for the tensorial, as opposed to vectorial, nature of distortions and culminate in (i)~distinct sum rules for odd responses, (ii)~correspondences between normal and shear stress responses, and (iii)~symmetry selection rules for response coefficients and their contrast with electromagnetic analogs, see SM~\cite{SI}. The nonmetricity and three-state QGTs capture the intrinsic geometry of two-band and three-band viscoelastic tensor contributions, as exemplified by the flat-band limit. As such, these constitute geometric higher-order analogs of the Berry curvature in the linear Hall viscosity of incompressible quantum Hall fluids~\cite{Avron1995}. Further, analogously to the linear viscous Hall responses, NOVEs dissipate no energy upon actions of strains and shears, as we demonstrate in SM~\cite{SI}.

We stress, however, that unlike the linear viscoelasticity, the nonlinear viscoelastic effects realize a~unique interplay of two-state and three-state contributions. In particular, the vanishing of the three-band term under different symmetries reflects the chirality of the electronic ground states supporting multiband NOVEs. These physical implications arise from the fact that the three-state tensors manifest in the skewness of the Wannier functions~\cite{Avdoshkin2025}.  Specifically, the three-band contribution $\eta^\mathrm{3\text{-band}}_{ij;kl,mn}$ is intrinsically three-dimensional, due to the symmetries of the three-state QGT, $Q_{ij;kl,mn}^{abc}$. Moreover, in this interpretation, we find that increasing the values of these QGTs, and thus the skewness~\cite{Avdoshkin2025} and chirality of the Wannier states, enables the distortions to couple more strongly, within a~direct proportionality to the higher geometric tensors.

We briefly comment on the implementation of distortions in theoretical contexts of nonlinear viscoelastic effects. For second-order responses, the adapted Peierls gauge formalism for stress tensors $T_{ij} = \delta H /\delta{w_{ij}} = -\frac{\sin(k_i a_0)}{a_0}\partial H /\partial{k_{j}}$~\cite{Shapourian2015
}, and elastic moduli tensors, is \mbox{justified} for small deformations $(|\mathfrak{u}|/a_0 \ll 1)$, consistently with the approach of Ref.~\cite{Shapourian2015}. We note that from a~theoretical angle, the formulation of the effect suggests that generalizations to higher dimensions, with $n$-band terms, provides a fruitful avenue for \mbox{further}~studies.

Finally, we address how to isolate the nonlinear viscoelastic effects experimentally. To realize NOVEs, a~protocol to generate shear forces should be adapted similarly to linear responses~\cite{Hoyos2012, Berdyugin2019}. However, in the presence of nontrivial $\eta_{ij;kl,mn}$ correlators, the scaling of the current response would be expected to deviate from linear. The degree of nonlinearity, combined with the transverse character of the response of systems subject to orthogonal distortions, e.g., $w_{xx}, w_{yy}$, should precisely reflect the qualitative phenomenological features of the odd effects identified in this work. Concerning material candidates, we anticipate that magnetic topological insulators, e.g., axion insulators, could naturally realize NOVEs, due to the three-dimensional multiband quantum geometry accessible in these phases. Consequently, effective $\theta$~vacua of axion insulators under realistic material setups \cite{Mong2010afm,Otrokov2019prediction,Jo2020afm} provide for a~promising route to not only generate, but also amplify NOVE responses.

In summary, we uncovered a class of exotic nonlinear odd viscoelastic effects. We showed that these effects naturally arise in magnetic three-dimensional topological phases, and also in topologically trivial phases with nontrivial three-dimensional and three-state quantum geometries. Hence, if measured in a bulk material, the response provides an experimental fingerprint for the intrinsic quantum state geometries of higher geometric tensors and unconventional topological invariants. Finally, our work on dc nonlinear viscoelastic effects paves the way for further studies of analogous time-dependent responses and resultant exotic phenomenologies in future theoretical and experimental pursuits.

\sect{Acknowledgments}The authors thank Joe Huxford and Giandomenico Palumbo for illuminating discussions. A.~J.~acknowledges funding from the School of Natural Sciences, University of Manchester. W.~J.~J.~acknowledges funding from the Rod Smallwood Studentship at Trinity College, Cambridge. M.~M.~was funded by an EPSRC ERC underwrite Grant No.~EP/X025829/1. R.-J.~S. acknowledges funding from an EPSRC ERC underwrite Grant No.~EP/X025829/1, and a Royal Society exchange Grant No. IES/R1/221060, as well as Trinity College, Cambridge. This research was supported in part by grant NSF PHY-2309135 to the Kavli Institute for Theoretical Physics~(KITP).

\bibliography{refs}

\section*{End Matter}

\sect{Quantum nonmetricity tensor}Here we detail the formal definition of the quantum nonmetricity \mbox{tensor}~\cite{Jain2026}. The nonmetricity is expressed as $N^{ba}_{ij,kl,mn} = -\Tilde{\nabla}_{mn}g^{ba}_{ij,kl}$, with $\tilde{\nabla}_{mn} = -\frac{\sin(k_m a_0)}{a_0}\tilde{\nabla}_{n}$, under the minimal coupling prescription introduced in the main text. Here, $\tilde{\nabla}_{k} \mathcal{O}_{ij} = \partial_{k} \mathcal{O}_{i j} - \tilde{\Gamma}_{l i k} \mathcal{O}{^{l}_{j}} -\tilde{\Gamma}_{l j k} \mathcal{O}{^{l}_{i}}$ is the covariant derivative with respect to the Hermitian connection $\tilde{\Gamma}_{ijk}^{ba}
= A_{i}^{ab} (\mathcal{D}_{j} A_{k})^{ba}$, $\mathcal{D}$ is the Berry covariant derivative defined as $\mathcal{D}_{i} O^{ab} = \partial_{i} O^{ab} - \text{i} [A_{i} + \mathcal{A}_{i},O]^{ab}$ for an arbitrary operator $O$~\cite{Liu2023, mehraeen2024quantum}, and $\mathcal{A}$ the band-diagonal Berry connection.

\sect{Elastic tensor forms}In the following, we provide the explicit expressions for the matrix elements of the stress tensor $T_{ij}$ and its higher order analogs $U_{ij,kl}$ and~$V_{ij,kl,mn}$, which read
\begin{align}
    T_{ij}^{ab} =& \delta^{ab} \partial_{ij} E^{b}_{\textbf{k}} + \text{i}~\Delta^{ab}A_{ij}^{ab}, \\
    U_{ij,kl}^{ab} =& \mathcal{D}_{ij} T_{kl}^{ab}, \nonumber \\
    =& \partial_{ij} T_{kl}^{ab} + \text{i}~T_{kl}^{ab} (\mathcal{A}_{ij}^b-\mathcal{A}_{ij}^a) \nonumber \\ &+ \text{i}\sum_{c \neq a,b}(T_{kl}^{ac}A_{ij}^{cb} - A_{ij}^{ac}T_{kl}^{cb}), \\
    V_{ij,kl,mn}^{ab} =& \mathcal{D}_{ij} U_{kl,mn}^{ab}, \nonumber \\
    =& \partial_{ij}U_{kl,mn}^{ab} + \text{i}~U_{kl,mn}^{ab} (\mathcal{A}_{ij}^b-\mathcal{A}_{ij}^a)  \nonumber \\ &+ \text{i}\sum_{c \neq a,b}(U_{kl,mn}^{ac}A_{ij}^{cb} - A_{ij}^{ac}U_{kl,mn}^{cb}),
\end{align}
where $\mathcal{D}_{ij} O^{ab} = \partial_{ij} O^{ab} - \text{i}~[A_{ij} + \mathcal{A}_{ij},O]^{ab}$ represents the distortion-space covariant derivative.

\sect{Details on models with NOVE} In the following, we further detail the constructions and exact parametrizations of model Hamiltonians realizing NOVEs. We consider a two-band Hopf insulator model that admits an adaptable tight-binding representation in real space~\cite{Moore2008}. This can be expressed as ${H_0 = \vec{d}\cdot\boldsymbol{\sigma}}$, with $d_i = \vec{z}^\dagger \sigma_i \vec{z}$ and a vector $\vec{z}$,
\begin{equation}
    \vec{z} = \begin{pmatrix}
    \sin (k_x a_0) + \text{i} \sin (k_y a_0)\\ \sin (k_z a_0) + \text{i} \left(\sum_i \cos (k_i a_0) - m \right)\end{pmatrix}.
\end{equation}
Here, $\vec{z}$ realizes nontrivial Hopf invariant $\chi=1$ for ${1<|m|<3}$ and $\chi=-2$ for $|m|<1$. 

For the three-band case, we begin with the model that possesses an invariant protected by chiral symmetry~\cite{Palumbo2019}. The Hamiltonian reads $H = \vec{d}\cdot \boldsymbol{\lambda}$, with $\vec{d} = [0,0,0,\sin (k_x a_0),\sin (k_y  a_0),\sin (k_z  a_0), m - \sum_i \cos (k_i a_0),0]^\text{T}$, where $i = x,y,z$, and $\boldsymbol{\lambda}$ is the eight-vector of Gell-Mann matrices that represents a basis of $3 \times 3$ Hermitian traceless matrices. In order to isolate the nonmetricity and three-state QGT contributions, we set the energies of this model explicitly to $(-1, 0 ,1)$. This flat-band limit causes the energy derivatives, and thus the metric- and curvature-dependent two-band terms, to~\mbox{vanish}. This model, however, still possesses a $\mathcal{M}_{xz}\mathcal{T}$ symmetry, which nullifies the response. To this end, keeping the bands flat, we add a three-band \mbox{perturbation}, 
\begin{equation}
 \Delta H = \delta \sin[(k_x + k_y)  a_0]\lambda_1, 
\end{equation}
with a perturbation strength $\delta$. The perturbation breaks the chiral symmetry and trivializes the chiral invariant $\nu_i$, but retains the nontrivial geometry of the unperturbed model.

\end{document}


\title{{\textsc{supplemental material}} \\ Nonlinear Odd Viscoelastic Effect}

\newcommand{\TCM}{{TCM Group, Cavendish Laboratory, University of Cambridge, J.\,J.\,Thomson Avenue, Cambridge CB3 0HE, UK}}
\newcommand{\UoM}{Department of Physics and Astronomy, University of Manchester, Oxford Road, Manchester M13 9PL, UK}
\newcommand{\KITP}{Kavli Institute for Theoretical Physics, University of California, Santa Barbara, CA 93106, USA}


\author{Ashwat Jain}
\email{ashwat.jain@postgrad.manchester.ac.uk}
\affiliation{\UoM}

\author{Wojciech J. Jankowski}
\email{wjj25@cam.ac.uk}
\affiliation{\TCM}
\affiliation{\KITP}

\author{M. Mehraeen}
\email{mandela.mehraeen@manchester.ac.uk}
\affiliation{\UoM}

\author{Robert-Jan Slager}
\email{robert-jan.slager@manchester.ac.uk}
\affiliation{\UoM}
\affiliation{\TCM}

\date{\today}

\maketitle

\tableofcontents 
\newpage

\section{Derivation of nonlinear odd viscoelastic responses}
We begin by deriving the general Kubo response to second order, performing a systematic expansion in perturbations and frequencies. For an observable $A(t)$, and for a system governed by a Hamiltonian $H = H_0 + H_1(t)$ with perturbation $H_1(t) = \sum_{i=1}^2 B_i(t) \phi_i(t)$, we write
\begin{equation}
\label{eq:expectation}
\begin{split}
    \langle A(t) \rangle &= \bra{\psi(t)} A \ket{\psi(t)} \\
    &= \bra{\psi_0} U^\dagger(t) A U(t) \ket{\psi_0}.
\end{split}
\end{equation}
Here we work in the Schr\"odinger representation with $U(t)$ the time evolution operator
\begin{equation}
    U(t) = \textit{T}~ \exp\left(-\text{i}\int_{-\infty}^t \mathrm{d}t' H_1(t') \right),
\end{equation}
and $\textit{T}$ the time ordering operator. Using the Dyson series truncated to second order
\begin{equation}
    U(t) \approx \mathsf{1} - \text{i} \int_{-\infty}^t \mathrm{d}t' H_1(t') - \int_{-\infty}^t \mathrm{d}t'\int_{-\infty}^{t'} \mathrm{d}t'' H_1(t')H_1(t''),
\end{equation}
we expand 
\begin{equation}
\label{eq:gen_resp}
    A(t) = \langle A(t) \rangle + \sum_{i=1}^2 \int_{-\infty}^\infty \mathrm{d}\tau_1~ \chi_{A;B_i}(t;\tau_1)\phi(t-\tau_1) + \sum_{i,j=1}^2\int_{-\infty}^\infty \mathrm{d}\tau_1\int_{-\infty}^\infty \mathrm{d}\tau_2~ \chi_{A;B_i,B_j}(t;\tau_1,\tau_2)\phi_i(t-\tau_1)\phi_j(t-\tau_2),
\end{equation}
where~\cite{Rostami2021}
\begin{align}
    \chi_{A;B_i}(t;\tau_1) &= \text{i}~ \Theta(\tau_1)\big\langle[B_i(t-\tau_1),A(t)]\big\rangle,\\
    \chi_{A;B_i,B_j}(t;\tau_1,\tau_2) &= \text{i}^2~\Theta(\tau_2-\tau_1)\Theta(\tau_1)\big\langle[B_j(t-\tau_2),[B_i(t-\tau_1),A(t)]]\big\rangle.
\end{align}
In Fourier space (Lehmann representation), $\chi_{A;B_1,B_2}$ reads ~\cite{Bhalla2022,Bhalla2024,Jain2026}
\begin{align}
\label{eq:gen_chi2}
    \chi_{A;B_1,B_2}(\omega_1^+, \omega_2^+) &= \sum_\mathcal{P} \sum_{a,b,c} \Bigg[\frac{T_{A}^{ab} T_{B_1}^{bc} T_{B_2}^{ca}}{\omega_1^+ + \omega_2^+ + \Delta^{ba}} \bigg ( \frac{f_{bc}}{\omega_1^+ + \Delta^{bc}}-\frac{f_{ca}}{\omega_2^+ + \Delta^{ca}}\bigg )-\bigg( V^{aa}_{A,B_1,B_2} f_a + f_{ab} \bigg[ \frac{T_{A}^{ab} U_{B_1,B_2}^{ba}}{\omega_1^+ + \Delta^{ab}} + \frac{1}{2} \frac{T_{B_1}^{ba} U_{A,B_2}^{ab} + T_{B_2}^{ba} U_{A,B_1}^{ab}}{\omega_1^+ + \omega_2^+ + \Delta^{ab}}\bigg]\bigg)\Bigg],
\end{align}
where 
    \begin{align}
        T_{A}
        &=
        \frac{\delta \mathcal{H}}{\delta A},
        \\
        U_{B_1,B_2}
        &=
        \frac{\delta^2 \mathcal{H}}
        {\delta A \;\delta B_1},
        \\
        V_{A,B_1,B_2}
        &=
        \frac{\delta^3 \mathcal{H}}
        {\delta A \; \delta B_1 \; \delta B_2},
    \end{align}
$\sum_{\mathcal{P}}$ denotes $(B_1, \omega_1) \leftrightarrow (B_2, \omega_2)$ internal permutation symmetry, $a,b,c$ are the Hamiltonian eigenstate (band) indices, $\omega_{1,2}^+ = \omega_{1,2}+\text{i}\epsilon_{1,2}$ are complex frequencies, the matrix element $\bra{a}O\ket{b}$ is written as $O^{ab}$  for brevity, ${\Delta^{ab} = E^a - E^b}$ are the band energy differences, ${f_{ab} = f_a - f_b}$ with $f_a = \mathrm{e}^{-\beta E^a}/\mathcal{Z}$, and finally $\mathcal{Z}$ is the partition function at temperature $1/\beta$. We emphasize here that $V_{A,B_1,B_2}$ is the nonlinear elastic modulus operator, which physically accounts for contact interactions at second order, in close analogy with electromagnetic response theory \cite{Parker2019}. Throughout this work, coordinate indices are denoted by $i,j,k,l,m,n$.

We can now define the second-order response tensor $\sigma_{A;B_1,B_2}$ as
\begin{equation}
\label{eq:sigma_from_chi}
    \sigma_{A;B_1,B_2} = \frac{-\text{i}\omega'^+}{i\omega_1^+ i\omega_2^+} \chi_{A;B_1,B_2}.
 \end{equation}
The frequency denominators are required when working in minimal coupling or velocity gauge, i.e., when the perturbation fields are gauge potentials~\cite{Bradlyn2012, Michishita2021}, while they do not appear while working in the length gauge~\cite{Parker2019}. The frequency numerator stems from the fact that $\chi_{A;B_1,B_2}$ as calculated above is a polarization susceptibility, as opposed to a current density response. Next, we~extract the odd (dissipationless) component of the response tensor $\sigma_{A;B_1,B_2}$ as~\cite{tsirkin2022}
\begin{equation}
    \sigma^\mathrm{odd} = \sigma^\mathrm{tot} - \sigma^\mathrm{diss},
\end{equation}
where the dissipative component of the response tensor is obtained on fully symmetrizing all indices. For linear response, this simply yields~\cite{tsirkin2022, Mandal2024}
\begin{equation}
    \sigma^\mathrm{odd}_{A;B_i} = \sigma^\mathrm{tot}_{A;B_i} - \sigma^\mathrm{diss}_{(A;B_i)} = \sigma_{[A;B_i]},
\end{equation}
i.e., the antisymmetrized component. Correspondingly, for the nonlinear case, we write
\begin{equation}
    \sigma^\mathrm{odd}_{A;B_1,B_2} = \sigma^\mathrm{tot}_{A;B_i,B_2} - \sigma^\mathrm{diss}_{(A;B_1,B_2)}. 
\end{equation}
The symmetry of the response functions in $B_1$ and $B_2$ allows simplification to
\begin{align}
    \sigma^\mathrm{odd}_{A;B_1,B_2} &= \sigma^\mathrm{tot}_{A;(B_i,B_2)} - \frac{1}{3}(\sigma^\mathrm{tot}_{A;(B_1,B_2)}+\sigma^\mathrm{tot}_{B_1;(A,B_2)}+\sigma^\mathrm{tot}_{B_2;(B_1,A)}) \nonumber \\
    &= \frac{1}{3}(2\sigma^\mathrm{tot}_{A;(B_1,B_2)} - \sigma^\mathrm{tot}_{B_1;(A,B_2)} -\sigma^\mathrm{tot}_{B_2;(B_1,A)}),\label{eq:symm_sigma}
\end{align}
where $\sigma_{\star;(\star,\star)}$ is the symmetrized form which uses $\chi_{A;B_1,B_2}$ as in Eq.~\eqref{eq:gen_chi2}. We consider the case of $n$ bands $\ket{1}, \ket{2} \ldots \ket{n}$ at zero temperature, restricting our attention to clean systems. To retrieve the dc response to purely static fields $\phi_i$, we take the symmetric dc limit $\omega_1 = \omega_2 \rightarrow 0$~\cite{Sipe2000}. The antisymmetric limit is more appropriate for optical responses, where it represents a rectification~\cite{Matsyshyn2019} to the zero-frequency divergences, and is related to the circular photogalvanic effect~\cite{deJuan2017}. In the symmetric frequency limit $\omega_1 = \omega_2 = \omega'/2 \equiv \omega$, the expression for the odd response reads
\begin{equation}
    \sigma^\mathrm{odd}_{A;B_1,B_2} = \lim_{\omega \rightarrow 0} \frac{2 \text{i}}{\omega^+} \bigg[
     \chi_{A;B_1,B_2}(2\omega,\omega,\omega) - 
     \chi_{B_1;A,B_2}(2\omega,\omega,\omega) - 
     \chi_{B_2;B_1,A}(2\omega,\omega,\omega)\bigg]. 
\end{equation}

Specializing to the case of elastic perturbations, to obtain the nonlinear elasticity tensor $\eta_{ij;kl,mn}$, we substitute ${A \rightarrow w_{ij}}$, ${B_1 \rightarrow w_{kl}}$, ${B_2 \rightarrow w_{mn}}$, such that $T_A \rightarrow T_{ij}, T_{B_1} \rightarrow T_{kl}, T_{B_2} \rightarrow T_{mn}$ (and similarly for $U$, $V$) are the stress and moduli \mbox{tensor} \mbox{operators} and $w_{ij} = \partial_i \mathfrak{u}_j$ is the distortion tensor. We implement strain by gauging momenta using a generalized Peierls \mbox{substitution}~\cite{Shapourian2015}
\begin{equation}\label{eq:peierls_gauge}
    k_j \rightarrow k_j - w_{ij} \frac{\sin(k_i a_0)}{a_0},
\end{equation}
with lattice constant $a_0$, which yields (denoting by $\partial_j$ the derivative with respect to $k_j$)
\begin{equation}
    T_{ij} = -\frac{\sin(k_i a_0)}{a_0} \partial_j H,\qquad U_{ij,kl} = \frac{\sin(k_i a_0) \sin{(k_k a_0)}}{a_0^2} \partial_j\partial_l H, \qquad V_{ij,kl,mn} = -\frac{\sin(k_i a_0)\sin(k_k a_0)\sin(k_m a_0)}{a_0^3} \partial_j\partial_l\partial_n H.
\end{equation}
Substituting into Eq.~\eqref{eq:gen_chi2}, we obtain a split into two- and three-band contributions. On further simplification, this yields~\cite{Jain2026}
\begin{align}
   \eta^\mathrm{2-band}_{ij;kl,mn}(\vec{k}) &= \frac{\sin(k_i a_0)\sin(k_k a_0)\sin(k_m a_0)}{a_0^3} \sum_{a,b}f_{ba}\bigg[ \frac{4 i \tau}{3} g^{ba}_{[(ln)}\partial_{j]}\Delta^{ba} + \frac{\Omega^{ba}_{j(l}\partial_{n)}\Delta^{ba}}{\Delta^{ba}} + \frac{32i}{3}N^{ba}_{[(ln)j]}\bigg],\label{eq:2band}\\
   \eta^\mathrm{3-band}_{ij;kl,mn}(\vec{k}) &= \frac{2}{3} \frac{\sin(k_i a_0)\sin(k_k a_0)\sin(k_m a_0)}{a_0^3} \sum\limits_{a,b,c}  \bigg(f_a \Delta^{bc}+f_b \Delta^{ca}+f_c \Delta^{ab}\bigg)\frac{\text{Re}Q_{j(ln)}^{abc}}{\Delta^{ba}}\bigg(2+\frac{(\Delta^{ba})^2}{\Delta^{bc}\Delta^{ca}}\bigg). \label{eq:3band}
\end{align}
Note that in the above, $\epsilon \rightarrow 1/\tau$, with $\tau$ a phenomenological relaxation time. These expressions can be recast in a more convenient form by defining the tensors
\begin{align}
A_{ij}^{ab} &\equiv \text{i}\bra{u^a}\ket{\partial_{w_{ij}}u^b},\\
    g_{ij,kl}^{ab} &\equiv \Re(A_{ij}^{ab}A_{kl}^{ba}),\\
    \Omega_{ij,kl}^{ab} &\equiv -2\Im(A_{ij}^{ab}A_{kl}^{ba}),\\
    N_{ij,kl,mn}^{ab} &\equiv -\tilde{\nabla}_{mn} g_{ij,kl}^{ab},\\
    Q_{ij;kl,mn}^{abc} &\equiv A_{ij}^{ab}A_{kl}^{bc}A_{mn}^{ca}.
\end{align}
Here, $\tilde{\nabla}_{mn}$ is the covariant derivative corresponding to the deformation-space Hermitian connection $\tilde{\Gamma}_{ij;kl,mn}^{ba}= A_{ij}^{ab} (\mathcal{D}_{kl} A_{mn})^{ba}$ with $\mathcal{D}_{ij} O^{ab} = \partial_{ij} O^{ab} - \text{i}~ [A_{ij} + \mathcal{A}_{ij},O]^{ab}$ the Berry covariant derivative~\cite{Jain2026}. These quantities can be thought of as dressed versions of their pure quantum geometric counterparts, where momentum derivatives have been replaced by distortion variations. Recasting in terms of these dressed geometric objects completes the derivation of Eqs.~(3)-(4) in the main text.

Note also that the symmetric (dissipative) part vanishes for both the two-band and three-band responses, resulting in the sum rule
\begin{equation}
\label{eq:sum_rule}
     \eta^\mathrm{odd,\,dc}_{ij;kl,mn} +  \eta^\mathrm{odd,\,dc}_{kl;mn,ij} +  \eta^\mathrm{odd,\,dc}_{mn;ij,kl} = 0,
\end{equation}
which shows explicitly that $\eta^\mathrm{odd,\,dc}_{ij;kl,mn}$ is the nondissipative response component.

Furthermore, this elucidates that all responses which put the three directions on equal footing (i.e., responses where $\{ij,kl,mn\} = \{xx,yy,zz\}$ or $\{xy,yz,zx$\}) must share the same form if their output index $j$ is the same, since in Eqs.~\eqref{eq:2band} and~\eqref{eq:3band}  the $\sin(k_x a_0)~\text{sin}(k_y a_0) \sin(k_z a_0)$ terms can be redistributed an arbitrary fashion. We thus get
\begin{equation}
    \eta_{xx;yy,zz}^\mathrm{odd, dc} = \eta_{yx;zy,xz}^\mathrm{odd, dc} = \eta_{zx;xy,yz}^\mathrm{odd, dc}.
\end{equation}
In addition, due to the construction in Eq.~\eqref{eq:symm_sigma}, the response is symmetric in last two index pairs, leading to six equal components of the tensor $\eta$. Similarly, the $yy;zz,xx$ and $zz;xx,yy$ responses each yield six equivalent $\eta$'s. Combined with the sum rule given by Eq.~\eqref{eq:sum_rule}, this implies that for arbitrary systems, there are two independent quadratic odd viscoelastic responses, for which we take the representatives $xx;yy,zz$ and $yy;zz,xx$. However, the presence of additional symmetries in certain systems may reduce the number of independent components of $\eta$ further to one independent component, as in the case of the Moore-Ran-Wen Hopf insulator model~\cite{Moore2008}, where $\eta_{xx;yy,zz} = \eta_{yy;zz,xx}$. The flattened chiral model considered also possesses such a symmetry, but it is lost upon adding the symmetry-breaking perturbation. We elaborate further on the role of symmetries in determining the response functions in the next Section.

\section{Symmetry analysis of viscoelastic response tensors}
The dissipationless elastic response tensor possesses a high degree of symmetry, which we elucidate below. In particular, we show that the presence of certain symmetries in the system cause some components of the bulk response to vanish, and provide a~compact set of selection rules in Table~\ref{tab:symmetries} which summarize the required conditions for a nonzero response. The symmetries under consideration are momentum-space mirror symmetry $\mathcal{M}_{xy}$, inversion symmetry $\mathcal{P}$, time reversal $\mathcal{T}$, and combinations thereof. These three elementary symmetries enforce the following relations on the Bloch functions $u$ and energies $\Delta$:
\begin{align}
    \mathcal{M}_{xy} &: u^a(k_x, k_y, k_z)=u(k_x, k_y,-k_z) \qquad\Delta(k_x, k_y, k_z) = \Delta(k_x, k_y, -k_z)\label{eq:symm_action_MT},\\\mathcal{P} &: u^a(\vec{k})=u^a(\vec{k})\qquad\qquad\qquad\quad\;\;\Delta(\vec{k}) = \Delta(-\vec{k}),\label{eq:symm_action_PT}\\
    \mathcal{T} &: u^a(\vec{k})=[u^a(-\vec{k})]^* \qquad\qquad\quad\;\;\;\Delta(\vec{k}) = \Delta(-\vec{k}).\label{eq:symm_action_T}
\end{align}
Here we take as a representative $\mathcal{M}_{xy}$ for concreteness, although the results apply equally well to $\mathcal{M}_{yz}$ and $\mathcal{M}_{zx}$. Note further that in describing the action of $\mathcal{T}$, we assume that the system is composed of spinless quasiparticles ($\mathcal{T}^2=+1$) and is fully nondegenerate so that the bands map onto themselves under the action of $\mathcal{T}$.

Using these relations, we find that all the terms in the response functions [Eqs.~(3) and~(4) of the main text] have well-defined parities under each of the symmetry operations, which are summarised in Table~\ref{tab:symmetries}. In particular, we note that the presence of a~combined momentum-space mirror and time-reversal symmetry $\mathcal{M}_{ij}\mathcal{T}$ causes all but the metric-dependent response component to vanish. Both models considered in the main text possess a definite parity under such a symmetry: the two-band MRW model~\cite{Moore2008} and the unperturbed three-band chiral model~\cite{Neupert2012,Palumbo2019} are even and odd respectively under $\mathcal{M}_{xz}\mathcal{T}$ and show only the metric-dependent response. We thus elicit the $\eta^{\text{2-band}}_{ij;kl,mn} (N)$ and $\eta^{\text{2-band}}_{ij;kl,mn} (Q)$ responses in the three-band chiral model by breaking the symmetry via a perturbation, and flattening the bands in order to nullify the $\eta^{\text{2-band}}_{ij;kl,mn} (g)$ and $\eta^{\text{2-band}}_{ij;kl,mn} (\Omega)$ responses which vary as the energy gradients.
\begin{table*}[t]
\centering
\begin{ruledtabular}
\begin{tabular}{c c c c c c c}
& $\mathcal{M}_{ij}$ & \textbf{$\mathcal{P}$} & \textbf{$\mathcal{T}$} & $\mathcal{M}_{ij}\mathcal{P}$ & $\mathcal{P}\mathcal{T}$ & $\mathcal{M}_{ij} \mathcal{T}$\\
\hline\\[-1ex]
$\eta^{\text{2-band}}_{ij;kl,mn} (g)$     & \ding{51}* & \ding{51}* & \ding{51}*  & \ding{51} & \ding{51} & \ding{51} \\[2ex]
$\eta^{\text{2-band}}_{ij;kl,mn} (\Omega)$& \ding{51}* & \ding{51}* & \ding{55}    & \ding{51} & \ding{55} & \ding{55} \\[2ex]
$\eta^{\text{2-band}}_{ij;kl,mn} (N)$     & \ding{51}* & \ding{51}* & \ding{55}    & \ding{51} & \ding{55}  & \ding{55} \\[2ex]
$\eta^{\text{3-band}}_{ij;kl,mn} (Q)$     & \ding{51}* & \ding{51}* & \ding{51}*  & \ding{51} & \ding{51} & \ding{55} \\[2ex]
\end{tabular}
\end{ruledtabular}
\caption{Selection rules showing the parities of various response coefficients in systems which are invariant under the action of momentum-space mirror symmetry $\mathcal{M}_{ij}$, inversion (parity) symmetry $\mathcal{P}$, time reversal symmetry $\mathcal{T}$ and their combinations $\mathcal{M}_{ij}\mathcal{P}$, $\mathcal{P}\mathcal{T}$ and $\mathcal{M}_{ij}\mathcal{T}$. The `\ding{55}' entries lead to a vanishing response on integration over the BZ, while the `\ding{51}' entries have a generically nonvanishing response. Furthermore, entries marked with a star ($*$) denote response coefficients which are nonzero in the viscoelastic response, but are expected to vanish in the corresponding electromagnetic case.}
\label{tab:symmetries}
\end{table*}
The presence of an additional index in the tensorial elastic terms allows control over the parity of the local response function, exemplified by the sine prefactor in the Peierls gauge substitution [Eq.~\eqref{eq:peierls_gauge}].

\section{Scaling of nonlinear viscoelastic effects in periodic Hamiltonians}
Here we show that the dissipationless Hall-like response coefficients for any $\vec{k}$-periodic Hamiltonian scale as
\begin{equation}
    \eta_{ij;kl,mn} \rightarrow p ~\eta_{ij;kl,mn}
\end{equation}
under the momentum scaling $k_z \rightarrow p~k_z$. Here, the choice of the $k_z$ direction is arbitrary and the arguments below apply equally well to scaling of other momentum components.

The Hall-like response coefficients we consider include all three orthogonal directions, i.e., $\{ij,kl,mn\} = \{xx,yy,zz\}$ or $\{ij,kl,mn\} = \{xy,yz,zx\}$. In particular, this implies that the second-order odd response functions, all of which possess $\partial_{ij}$, $\partial_{kl}$ and $\partial_{mn}$ derivatives, must have exactly one derivative in each momentum coordinate after the Peierls substitution [Eq.~\eqref{eq:peierls_gauge}]. The $k_z$ derivative induces an overall multiplicative factor of $p$ as a result of the $k_z \rightarrow p~k_z$ scaling, while the entirety of the remaining expression is $2\pi$-periodic as a result of the periodicity of the Hamiltonian. We can thus write
\begin{equation}
    \eta_{ij;kl,mn}^{(p)} (k_x, k_y, k_z) = p J(k_x, k_y, pk_z),
\end{equation}
where $J$ is a $2\pi$-periodic function. On integration over the BZ, we see that 
\begin{align}
    \eta_{ij;kl,mn}^{(p)} &= p \int_{-\pi}^{\pi}\mathrm{d}k_x \int_{-\pi}^{\pi}\mathrm{d}k_y \int_{-\pi}^{\pi} \mathrm{d}k_z J(k_x,k_y,pk_z) \nonumber \\
    &= \int_{-\pi}^{\pi}\mathrm{d}k_x \int_{-\pi}^{\pi}\mathrm{d}k_y \int_{-p\pi}^{p\pi}\mathrm{d}k_z' J(k_x,k_y,k_z') \nonumber \\
    &= p \int_{-\pi}^{\pi}\mathrm{d}k_x \int_{-\pi}^{\pi}\mathrm{d}k_y \int_{-\pi}^{\pi} \mathrm{d}k_z' J(k_x,k_y,k_z') \nonumber  \\
    &= p \int_{-\pi}^{\pi}\mathrm{d}k_x \int_{-\pi}^{\pi}\mathrm{d}k_y \int_{-\pi}^{\pi} \mathrm{d}k_z J(k_x,k_y,k_z) \nonumber  \\
    &= p~ \eta_{ij;kl,mn}^{(1)},
\end{align}
where $k_z'$ = $p k_z$ and we have used the periodicity of $J$ in going from the second to the third line. This shows that the nonlinear odd viscoelastic responses scale exactly with integer scaling of the momenta.

\section{Numerical details}
In this Section, we provide details of the numerical methods used to obtain Fig.~2 and Fig.~3 in the main text. First, a numerical estimate of the Hopf invariant for the MRW model was found using the construction presented in Ref.~\cite{Jankowski2025Gerbes}, as well as by numerically finding the preimage of two antipodal points on the two-sphere $(S^2)$ under the Hopf map $\textbf{d}(\vec{k})$ defined via
\begin{equation}
    H(\vec{k}) = \textbf{d}(\vec{k})\cdot\boldsymbol{\sigma},
\end{equation} where $\boldsymbol{\sigma}$ is the vector of Pauli matrices. Both methods show agreement about the value of the Hopf invariant and its dependence on the variation of the mass parameter $m$ as well as the momentum scaling $p$.

All integrals for Fig.~3 in the main text were performed numerically over a discretized $\textbf{k}$-grid. The presence of the Hopf invariant leads to an obstruction to a globally smooth periodic gauge, and Fig.~3 in the main text  was obtained by computing Wannier functions in a maximally smooth (but nonperiodic) gauge for every value of mass parameter~$m$. For small distortions, the position of the Wannier center was tracked with the derivative taken using a centered finite-difference over the distortion points.

\newpage

\bibliography{refs}